\documentstyle[11pt]{article}
 
\begin{document}

\begin{center}

S-TREE ANALYSIS OF THE SUBSTRUCTURE OF\\
THE COMA CLUSTER
\vspace{0.2in}

V.G.Gurzadyan$^{1,2}$ and A.Mazure$^3$
\vspace{0.1in}

1. Sussex University, Brighton, UK;

2. Yerevan Physics Institute, Armenia

3. Laboratoire d'Astronomie Spatiale, Marseille, France
\vspace{0.2in}

\end{center}

The S-tree method is used to estimate the redshift of the core of Coma
cluster - $cz = 6953$ km s$^{-1}$ and dispersion $\sigma=949$ km s$^{-1}$.
The existence of three subgroups of galaxies is revealed,
one of them is associated with the cD galaxy NGC 4874, another -
with NGC 4889.
It is argued that these subgroups are 'galaxy associations', i.e.
galactic dynamical entities moving through the main cluster. The
non-stationarity of the dynamical processes ongoing in the Coma core
is concluded.

\vspace{0.2in}

The structure of the Coma cluster
was attracting much attention during decades 
(see Biviano's talk at this conference for a review).
 These
studies on one hand involve various statistical tools of the analysis of
observational data, the wavelets being among the most recent ones, and
creation of various models based on certain assumptions on the cluster's
symmetry, equilibrium, dark matter distribution, etc., on the other hand.

The X-ray observations of the Coma cluster, especially by
ROSAT, add crucial information
concerning the substructures, as well as the processes governing the
mechanisms of X-ray emission. The further combination of data on the
galaxies and X-ray data will enable much deeper insight on the structure
of the cluster.

In the present study we use S-tree method to analyze the core substructure
of the Coma cluster.
The S-tree technique (Gurzadyan \& Kocharyan 1994;
 Bekarian \& Melkonian 1997) is based on the powerful
methods of theory of dynamical
systems and already was used to study the substructure of
the Local Group, Virgo cluster and of a sample of ENACS Abell clusters
(see Gurzadyan \& Mazure 1996, 1997 and references therein).

{\bf Data}. We have used the data compiled by Biviano et al (1996); their
dataset is based
on their own observations with CFHT, the data by Colless and Dunn (1996)
and those obtained from literature.
 From these dataset we have taken the sample of 188
galaxies of $3000''\times 3000''$ field centred
on $\alpha=12^h 57^m.3$, $\delta=28^{\circ} 14'.4$ brighter than $18^m.0$.
Our choice was determined by considerations on data completeness with the
magnitude.

{\bf Method.} The S-tree method is using the information on the
2D coordinates, redshifts and magnitudes of galaxies in a self-consistent
way, i.e. revealing the correlation which should exist between the
parameters of a gravitationally interacting N-body system. It is done via
following the properties of the flow of geodesics in phase space, so that
the so-called two-dimensional curvature
of the phase space of the system
$K_{\mu\nu}=R_{\mu\nu\rho\delta}u^{\rho}u^{\delta}$ ($R$ is the Riemann
tensor, $u$ is the velocity of geodesics) is used for the
evaluation of the 'degree
of boundness'; for details see (Gurzadyan \&
Kocharyan 1994). This procedure enables to reveal the
hierarchical structure of the system including the existence of subgroups
and representation of the result via well known form of
tree-diagrams.

{\bf Results.} The S-tree had revealed the following substructure in the
core of Coma cluster: the main system (MS) containing 174 galaxies centred
on $\alpha=12^h 57^m 32^s.3$, $\delta=28^{\circ} 19'.31^"$ and 3
subgroups: of 34 galaxies (1s), 14 galaxies (2s) and 16 galaxies (3s).
The 1st subgroup contains the 2nd brightest galaxy of the Coma core - NGC
4874. The centre of this subgroup is at $\alpha=12^h 57^m 34^s.31$,
$\delta=28^{\circ} 15' 35^".45$, i.e. is not coinciding with NGC 4874.
 The obtained parameters of the galaxies of the MS and the subgroups are given
in the Table 1, which includes the number of galaxies (N), the median
velocity (m, in km s$^{-1})$), standard deviation of the redshift
distribution ($\sigma$, in km s$^{-1}$),
3rd and 4th moment of redshift distribution, (s) and (c), respectively.

\begin{table*}
\centering
\caption{Parameters of the Coma core main system (MS) and subgroups (1s,
2s, 3s): N denotes the number
of galaxies in the initial sample (T) and in each system; m the median
velocity; $\sigma, s, c$, the standard deviation, 3rd and 4th moment of
redshift distribution, respectively.}
\medskip
\begin{tabular}{lllllll}
\hline
\hline
Coma core    & T ($<18m$)& MS      & 1s       &  2s     & 3s            &
\\
\hline
N            & 188       & 174     & 34       &  14      & 16           & \\
m            &           & 6953    & 6892     & 7563     & 6013         & \\
$\sigma$     &           & 949     & 206      &  60      & 122          & \\
s            &           & -0.2    & 0.4      & 0.4      & 0.1          & \\
c            &           & -0.86   &-1.1      &-1.0      &-1.4          & \\
\hline
\end{tabular}
\end{table*}

{\bf Discussion}.

The results obtained above
enable one to draw the following picture on the substructure and the dynamical
processes evolving in the Coma core.

First, via S-tree we have obtained the redshift and the velocity
dispersion of the main body of Coma cluster: $cz = 6953$ km s$^{-1}$ and
$\sigma=949$ km s$^{-1}$,
respectively. These parameters do not differ much from those obtained before
(Biviano et al, 1996, Colless \& Dunn 1996), except the centre of the cluster
does not coincide exactly with the dominant galaxy NGC 4874, though lies in
its vicinity.

We also note that though there is some overlapping in the redshift
distribution of the substructures, nevertheless the three
subgroups are well separated in the redshift space. This can be
interpreted as a result of essential mutual bulk motion of subgroups,
i.e. when
the bulk velocity is exceeding the velocity dispersion of each subgroup.
If so, this indicates the ongoing merging of the subgroups.

The fact that NGC 4874 ($\alpha=12^h 57^m 27^s.38$,
$\delta=28^{\circ} 13' 43^"$, $cz=7176 km$ $s^{-1}$)
is not situated in the mass centre of subgroup 1
and its redshift does not coincide with the median redshift of the subgroup,
and also the members of the subgroup are mainly bright galaxies,
also shows that this subgroup is moving through the cluster, resulting
a redistribution of galaxies due to dynamical friction.
Biviano et al (1996) proposed to explain the overdensity of bright galaxies
in the vicinity of NGC 4874 by the core-halo segregation mechanism
during the own evolution of the subgroup. However, this mechanism
is efficient in stellar systems with large number of stars, while
its characteristic time scales are too large for the group of galaxies.
The dynamical friction, on the other hand, which
does depend on the parameters of the moving objects, can be responsible
for the observed segregation.
Subgroup 1 shows better separation by 'degree of boundness' from the
galaxies of the main cluster in redshift space, while the galaxies of subgroup
3 are more overlapped with redshifts of main cluster, i.e. there are galaxies
not belonging to subgroup 3, but having redshifts lying within the
redshift interval of that subgroup.
This is indicating the ongoing dissolution of subgroup 3 within the main
cluster, which is then an elder merger; this conclusion is supported with the
essential shift of the velocity of NGC 4889
($\alpha=12^h 56^m 55^s$,
$\delta=28^{\circ} 14' 46^"$, $cz=6497 km$ $s^{-1}$), from the median
velocity of subgroup 3.
Similar conclusion is has been drawn by Colless and Dunn (1996) from other
considerations.

The recent S-tree analysis of the substructure of a sample of ENACS clusters
(Gurzadyan \& Mazure 1997) enables to
claim the existence of {\it galaxy associations} in the clusters, i.e.
galactic dynamical entities with truncated velocity distributions. It was 
shown that
the motion within the main cluster can be the natural mechanism explaining
such truncation.  The present study reveals the existence of such
subgroups - galaxy associations -
in Coma cluster which are undergoing the merging process but are
in various phases of the merging. Indeed, the truncation of redshift
distribution for the elder merger
(subgroup 3) is more evident than for subgroup 1, moreover only
a core of galaxies had survived in the dissolved subgroup 3.

These results enable us to draw the basic conclusion on the {\it
ongoing non-stationary dynamical processes in the core of Coma cluster}.

This conclusion has to affect the interpretation of the X-ray
data. Particularly, the isothermal assumption of the X-ray gas has to be
considered as too simplified, otherwise it will lead to overestimation of
the mass. The multi-temperature gas will mean that the X-ray flux peaks
will depend on the wavelength, and therefore the correlation
of the X-ray peaks and galaxy distributions may be rather complicated.

The properties of bulk flows of subgroups
is another key for the understanding of the non-stationary
processes occuring in Coma and other clusters of galaxies
(Gurzadyan \& Rauzy, 1997).

V.G. thanks J.Barrow and the staff of Sussex Astronomy Centre for
hospitality; while in Sussex V.G. was supported by Royal Society.


\begin{thebibliography}{}


\bibitem{Bi2}
Biviano A., 1997, this Proceedings.

\bibitem{BM}
Bekarian K., Melkonian A. 1997, Astrofizika, 40, 425.

\bibitem{Bi}
Biviano A., et al 1996, A\&A, 311, 95.

\bibitem{CD}
Colless M., Dunn  1996, ApJ, 458, 435.

\bibitem{GK1}
Gurzadyan V.G., Kocharyan A.A. 1994, Paradigms of the Large-Scale Universe,
  Gordon and Breach.

\bibitem{GM}
Gurzadyan V.G., Mazure A. 1996, Observatory 116, 391.

\bibitem{GM}
Gurzadyan V.G., Mazure A. 1997, MNRAS (in press); astro-ph/9709210.

\bibitem{GR}
Gurzadyan V.G., Rauzy S. 1997, Astrofizika 40, 473; astro-ph/9707198.

\end{thebibliography}
\end{document}